%% file: main.tex
\documentclass[aps,prb,reprint,floatfix]{revtex4-2}
\usepackage{graphicx}
\usepackage{amsmath}
\usepackage{amssymb}
\usepackage{physics}
\usepackage{subfigure}
\usepackage{orcidlink}

\usepackage{hyperref} 
\hypersetup{
	citecolor = blue,
	colorlinks = true,
	urlcolor = blue
}

\newcommand{\al}{\alpha}
\newcommand{\e}{\epsilon}
\newcommand{\si}{\sigma}
\newcommand{\la}{\lambda}
\newcommand{\p}{\prime}

\newcommand{\up}{\uparrow}
\newcommand{\down}{\downarrow}

\newcommand{\lb}{\left(}
\newcommand{\rb}{\right)}
\newcommand{\lB}{\left\{}
\newcommand{\rB}{\right\}}
\newcommand{\LB}{\left[}
\newcommand{\RB}{\right]}

\newcommand{\vv}[1]{\boldsymbol{#1}}

\begin{document}
\date{May 6, 2026}
\title{Nonlocal transport phenomena in coupled quasiperiodic Kitaev chains}
\author{Koki Mizuno\, \orcidlink{0009-0004-8276-1045}}
\email{mizuno.koki.t8@s.mail.nagoya-u.ac.jp}
\affiliation{Department of Physics, Nagoya University, Nagoya 464-8602, Japan}

\begin{abstract}
    We investigate the topological phases in a coupled one-dimensional p-wave superconducting Fibonacci quasicrystal modeled by the quasiperiodic Kitaev chain.
Recent studies have shown that the coupled system can host topological edge modes with Majorana fermions and enhance their topological protection, depending on the pattern of quasiperiodicity.
In this work, we elucidate the topological phases of the coupled system and demonstrate the dependence of differential conductance on the lead connecting pattern employing the Keldysh formalism and the recursive Green's function method.
Our findings reveal the emergence of topological phases in the coupled system, which are characterized by the presence of Majorana edge modes and the seepage of the Majorana wave function.
Furthermore, we identify a new topological phase transition induced by quasiperiodicity in the coupled system.
\end{abstract}

\maketitle

\section{Introduction}
\input{Intro}

\section{Formulations}\label{sec_Form}
\input{Formulations}

\section{Numerical calculation}\label{sec_Nume}

\input{Numerical}

\section{Topological invariant}\label{sec_invariant}
\input{Invariant}

\section{Topological phase transitions of coupled system}\label{sec_phase}
\input{Phase}

\section{Equivalence to \texorpdfstring{$s$-wave}{s-wave} proximity-induced system}\label{sec_Equv}
\input{equivalence}

\section{Conclusion and discussion}
In this paper, we clarified the properties of the topological phase in the coupled Kitaev chain system and found the role of quasiperiodicity in the topological phase transition.
To define the topological phase, we used the pseudospectral method. 
This method defines the topological invariant in real space with open boundary conditions, which also applies to the quasiperiodic system.
In the coupled system, we could see three values of topological invariant $\nu = 0$, $1$, and $2$, where $\nu = 0$ is a trivial phase, and the latter two are non-trivial phases.
Furthermore, the non-trivial phase with $\nu = 1$ is separated into two different types of phases according to the zero-bias value of the local and nonlocal conductance, namely, details of the Majorana edge modes.
In this paper, we have investigated the formulations of differential conductance of a coupled system connected to two half-infinite leads using Keldysh Green's function approach. 
By using this approach, we found the nonsymmetric Majorana wave function seepage at the edge of the coupled system in the $\nu = 1$ phase.
In one of the $\nu =1 $ phases, the Majorana wave function seeps into one of the coupled chains at the left edge, and in another, it seeps into another of the coupled chains at the right edge.

The above are the common properties between periodic and quasiperiodic systems.
In this study, we found new topological phase transitions as the effect of quasiperiodicity.
If the coherence length is long enough, the new topological phase transitions appear in non-trivial phases in the periodic system.
This can be intuitively understood as likely that effects of quasiperiodicity appear when the coherence length is long enough for Cooper pairs to sense quasiperiodicity.
Therefore, for the approximants, we can expect that the ratio between the coherence length and the unit cell size determines whether topological phases induced by quasiperiodicity exist.
Future research will examine the specific behavior of topological phases in approximants, which is considered to be strongly dependent on the model and system details.

Through calculations of differential conductance, we find that the zero-bias conductance peaks are nearly unchanged by quasiperiodicity in the parameter regimes studied here.
These results suggest that the transport signatures associated with localized Majorana modes are insensitive to the quasiperiodic modulation considered in this work.
Conversely, the topological phases of the coupled system and the nonsymmetric seepage of Majorana fermions appear in both periodic and quasiperiodic coupled chains in the parameter regimes studied here.

Our calculations of the conductance are easily expandable to two-dimensional and some three-dimensional systems.
However, we must investigate the calculation method to replace the Bloch theorem to consider a more complicated situation.
Therefore, in future work, we must construct the theory without relying on translational symmetry to analyze the quantum transport phenomena in quasicrystalline superconductor \cite{deguchi2012quantum,deguchi2015superconductivity,kamiya2018discovery, Shiino_SC_2021, wang2024superconductivity_QC}, ferromagnet \cite{tamura2021experimental,sato2017quantum}, antiferromagnet \cite{Ishikawa_AFM2018} and so on.
Finally, our discovery about the existence and isolation of Majorana fermions at the edge of coupled systems are expected to be verified in the systems of quasiperiodic quantum wire with $s$-wave proximity effect or can be simulated by the coupling of isotropic and anisotropic XY spin systems.

\begin{acknowledgments}
The author is deeply grateful to Masahiro Hori for his invaluable guidance, extensive discussions, and generous intellectual support throughout the development of this work.
This work is supported by JST SPRING (Grant No. JPMJSP2125).
\end{acknowledgments}

\section*{Data Availability Statement}
The data that support the findings of this article are available from the author upon reasonable request.

\appendix 
\section{Single chain system}\label{sec_App_single}
\input{Appendix/Single-chain}

\section{Derivation of the differential conductance in coupled system}\label{sec_App_A}
\input{Appendix/Formulation.tex}

\bibliography{ref}

\clearpage
\end{document}

%% file: Intro.tex
Since topological superconductivity has been advocated, research on the Majorana fermions emergence in materials has become one of the hottest topics in condensed matters physics \cite{Sato_Ando_2017, Leijnse_Flensberg_2012, Sato_Fujimoto_2009}.
Due to their non-Abelian statistics, the Majorana fermions in materials are expected to apply to quantum computing technologies \cite{lutchyn2018majorana, Chetan_2008, Majorana_wire_comp, Sarma_Freedman_Nayak_2015}, and have been extensively studied both experimentally and theoretically \cite{Zhang_2018, Sasaki_2011, Mandal_2023}.
A prototypical theoretical model for realizing Majorana fermions as an edge mode is the Kitaev chain \cite{Kitaev_2001}, which describes a spinless one-dimensional p-wave superconductor.
The Kitaev chain can be effectively realized in a quantum wire with spin-orbit coupling, $s$-wave superconducting proximity effect, and a Zeeman field \cite{Oreg_2010, Lutchyn_2010, Sato_Takahashi_Fujimoto_2010}.
Experimentally, several groups have reported signatures suggesting the Majorana edge modes in such systems \cite{Deng_Yu_Huang_Larsson_Caroff_Xu_2012,rokhinson2012fractional,das2012zero,albrecht2016exponential,deng2016majorana}.
One of the key signatures of Majorana fermions is the appearance of a zero-bias conductance peak of the differential conductance, quantized in units of $e^2/h$ \cite{Takagi_Tamura_2020, Linear_nonLinear_Kitaev, Pienrka_2012, KTLaw_2009, Asano_Tanaka_2013}.

In recent years, it has been reported that quasiperiodicity enhances the robustness of topologically protected Majorana edge modes in coupled quasiperiodic Kitaev chains \cite{Kitaev_stack}.
This finding suggests that quasiperiodicity can give rise to novel physical phenomena likely for many other systems, including the Kitaev chain, such as phase transitions, time-quasiperiodic behavior, and other emergent properties \cite{ghadimi2017majorana, Roy_phase_transition_qp, time_quasiperiodic, jiang_nonHermitian, topo_local_Kitaev, topo_critical_quasicrystal, Fibonacci_SC_PRB, Fibonacci_SC_arxiv, josephson-effect-Fibonacci, Three-TC-Fibonacci,yang2025emergent}.
Motivated by these findings, we investigate the nonlocal differential conductance and topological phase transitions of the coupled system.
The nonlocal conductance is formulated using the real-space Keldysh Green's function.
To identify topological phase transitions, we employ a topological index defined via the pseudospectral method \cite{K-theory_pseudo,saito2024, Fulga-topo-SC}.
In this study, we address two key questions:
(1) How does quasiperiodicity affect the nonlocal differential conductance?
(2) Does the coupled Kitaev chain model exhibit characteristic phenomena induced by quasiperiodicity?

Our findings indicate that, for the parameter regimes examined here, the answer to the first question is largely negative.
The differential conductance behaves nearly identically in periodic and quasiperiodic systems, suggesting that quasiperiodicity does not significantly alter transport signatures associated with localized Majorana modes in these regimes.
In contrast, the answer to the second question is affirmative.
We identify a novel topological phase transition induced by quasiperiodicity when the coherence length is sufficiently long to capture its effects.

Furthermore, in periodic and quasiperiodic coupled systems, we find topological phases that cannot be fully characterized by the topological index alone.
These phases exhibit distinct properties of the Majorana fermions at the edges, particularly an asymmetric wavefunction seepage.
For instance, a Majorana fermion may be delocalized between the coupled chains at the right edge, while at the left edge, a single Majorana fermion remains localized within only one of the coupled chains.
Finally, we establish an equivalence between the quasiperiodic Kitaev chain and an $s$-wave proximity-induced quasiperiodic quantum wire with strong spin-orbit coupling and a Zeeman field orthogonal to the wire.
This equivalence suggests our predictions can be experimentally verified without realizing a quasiperiodic one-dimensional $p$-wave superconductor.

This paper is organized as follows.
In Sec.~\ref{sec_Form}, we derive the differential conductance for a coupled system by Keldysh Green's function method.
In the next section, Sec.~\ref{sec_Nume}, we will show the results of numerical calculations of differential conductance.
Next, we define the topological index for the system with the pseudospectral method in Sec.~\ref{sec_invariant}.
According to this topological index and the zero-bias conductance value, we will see the topological phase transition and the effect of quasiperiodicity of the coupled system in Sec.~\ref{sec_phase}.
In the last of this paper, Sec.~\ref{sec_Equv}, we claim the quasiperiodic Kitaev chain can be effectively realized using the quasiperiodic quantum wire and $s$-wave superconductor.

%% file: Formulations.tex
\begin{figure}
    \centering
    \includegraphics[width=\columnwidth]{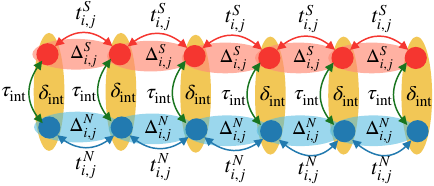}
    \caption{Schematic drawing of the coupled system.
    The arrows represent the hopping between the nearest-neighbor and interchain sites in this figure.
    The ellipses represent the coupling with the nearest-neighbor and interchain nearest site.
    If this system has quasiperiodicity, then nearest-neighbor hopping and coupling depend on the distance between neighbor sites.
    }
    \label{Fig_stack}
\end{figure}

This section formulates the differential conductance of a coupled system consisting of normal and superconducting Kitaev chains, as illustrated in Fig.~\ref{Fig_stack}.
This model was originally proposed in Ref.~\cite{Kitaev_stack}.
In the figure, we denote the hopping and pair potential in the superconducting chain as $t^{\rm S}_{i,j}$ and $\Delta^{\rm S}_{i,j}$, respectively.
Similarly, those in the normal chain are denoted as $t^{\rm N}_{i,j}$ and $\Delta^{\rm N}_{i,j}$.
Additionally, we introduce the interchain hopping and pair potential, denoted as $\tau_{\rm int}$ and $\delta_{\rm int}$.
To formulate the differential conductance, we follow a parallel discussion to that of the single-chain model presented in Ref.~\cite{Linear_nonLinear_Kitaev}.
For completeness, a brief review of the differential conductance in a single-chain model is provided in Appendix \ref{sec_App_single}.
For the coupled Fibonacci Kitaev chains with $N$ sites, the Bogoliubov–de Gennes (BdG) Hamiltonian matrix is represented by a $4N\times 4N$ matrix, given by
\begin{align}
    H_{S/N} &= \sum_{i=1}^{N}\LB \mu^{\rm S/N} d_{i}^{\rm S/N,\dag} d_{i}^{\rm S/N}\RB 
    \notag\\
    +&\sum_{i=1}^{N-1}\LB -t^{\rm S/N}_{i,i+1} d_{i+1}^{S/N\dag} d_{i}^{\rm S/N} +  h.c. \RB
    \notag\\
    +&\sum_{i=1}^{N-1}\LB \Delta^{\rm S/N}_{i,i+1}d_{i}^{\rm S/N}d_{i+1}^{\rm S/N}  + h.c. \RB,
    \label{Ham_stack_SN}
\end{align}
\begin{align}
    H_{\rm int} &=
    \sum_{i=1}^{N}\LB
    -\tau_{\rm int} d_{i}^{S\dag}d_{i}^{\rm N} 
    + \delta_{\rm int} d_{i}^{\rm S}d_{i}^{\rm N} + h.c.
    \RB,
    \label{Ham_stack_int}
\end{align}
where Hamiltonian (\ref{Ham_stack_SN}) describes the superconducting and normal Kitaev chains, and Hamiltonian (\ref{Ham_stack_int}) describes the interchain coupling.
The parameters $\mu^{\rm S}$ and $\mu^{\rm N}$ in Hamiltonian (\ref{Ham_stack_int}) are the chemical potential of the superconducting and normal Kitaev chains, respectively.
According to Ref.~\cite{Kitaev_stack}, this system hosts Majorana modes at the edges of both the normal and superconducting Kitaev chains, with their stability being enhanced by quasiperiodicity.
To investigate this, we calculate the nonlocal differential conductance of the system.
We consider the following cases;
\begin{itemize}
    \item[\textit{1}.] $S_{L/R}-S_{R/L}$ channel 
    \item[\textit{2}.] $N_{L/R}-N_{R/L}$ channel 
    \item[\textit{3}.] $S_{L/R}-N_{R/L}$ channel 
    \item[\textit{4}.] $S_{L/R}-N_{L/R}$ channel 
\end{itemize}
In these four cases, for example, Case \textit{3} corresponds to the scenario where one lead is connected to the left (right) edge of the superconducting Kitaev chain, while the other lead is connected to the right (left) edge of the normal Kitaev chain.

To derive the differential conductance, we define the matrix representation of the above Hamiltonian in real space basis as
\begin{equation}
    \begin{split}
        H_{\rm{couple}} =& \mqty(d^{S\dag} & d^{\rm S} & d^{N\dag} & d^{\rm N} ) \widetilde{H}_{\rm{couple}} \mqty(d^{\rm S} \\ d^{S\dag} \\ d^{\rm N} \\ d^{N\dag} ) ,
        \\
        \check{H}_{\rm{couple}} =&
        \mqty(
        \widetilde{H}_{\rm S}                 &  \widetilde{H}_{\rm{int}} \\
        \widetilde{H}_{\rm{int}}^{\dag} &  \widetilde{H}_{\rm N}
        ),
        \\
        \widetilde{H}_{S/N} =& \mqty(
        \widehat{\mu}^{\rm S/N}              & \widehat{\beta}_{\Delta}^{\rm S/N} &                        & \\
        \widehat{\beta}_{\Delta}^{S/N\dag} & \ddots            & \ddots                 & \\
                               & \ddots            & \ddots                 & \widehat{\beta}_{\Delta}^{\rm S/N} \\
                               &                   & \widehat{\beta}_{\Delta}^{S/N\dag} & \widehat{\mu}^{\rm S/N}  
        ) ,
        \\
        \widehat{\beta}_{\Delta}^{\rm S/N} =& \mqty(
        -t_{i,i+1}^{\rm S/N}        & -\Delta_{i,i+1}^{\rm S/N} \\
        \Delta_{i,i+1}^{\rm S/N}    & t_{i,i+1}^{\rm S/N}
        ),
        \\
        \widehat{\mu}^{\rm S/N}
        =&
        \mqty(
        -\mu^{\rm S/N} & 0\\
        0 & \mu^{\rm S/N}
        ),
        \\
        \widetilde{H}_{\rm{int}} =& \mqty(
        -t_{\rm int} I_{\rm N}  &  -\Delta_{\rm int} I_{\rm N} \\
        \Delta_{\rm int} I_{\rm N} & t_{\rm int} I_{\rm N}
        ).\label{Ham_Stack_mat}
    \end{split}
\end{equation}

In the last of this section, we derive the formulations of differential conductance for the four channels.
To facilitate the following discussion, we indicate the self-energy due to connecting with lead at the edge.
According to the basis in Eq.~(\ref{Ham_Stack_mat}), the self-energies are written as follows
\begin{align}
    \check{\Sigma}^{\rm S}_{L/R} =\mqty(
    \widetilde{\Sigma}_{L/R} & \widetilde{O}_{2N} \\
    \widetilde{O}_{2N}       & \widetilde{O}_{2N}
    ),
    \quad
    \check{\Sigma}^{\rm N}_{L/R} =\mqty(
    \widetilde{O}_{2N} & \widetilde{O}_{2N} \\
    \widetilde{O}_{2N} & \widetilde{\Sigma}_{L/R}
    ),\label{self_ene_4N}
\end{align}
where $\widetilde{\Sigma}_{L/R}$ are defined as
\begin{align}
    \widetilde{\Sigma}_{\rm L}
    &= \mqty[
    \widehat{\Sigma}_{\rm L} & \widehat{O}_2  & \cdots & \widehat{O}_2 \\
    \widehat{O}_2  & \widehat{O}_2  & \cdots & \widehat{O}_2 \\
    \vdots     & \vdots     & \ddots & \vdots \\
    \widehat{O}_2  & \widehat{O}_2  & \cdots & \widehat{O}_2 
    ], 
    \\
    \widetilde{\Sigma}_{\rm R}
    &= \mqty[
    \widehat{O}_2  & \cdots & \widehat{O}_2  & \widehat{O}_2 \\
    \vdots     & \ddots & \vdots     & \vdots \\
    \widehat{O}_2  & \cdots & \widehat{O}_2  & \widehat{O}_2 \\
    \widehat{O}_2  & \cdots & \widehat{O}_2  & \widehat{\Sigma}_{\rm R}
    ],\label{self-ene-2N}
\end{align}
where $\widehat{O}_2$ is the $2\times 2$ zero matrix.
The self-energies $\widehat{\Sigma}_{\rm L/R}$ are defined by using the Green's function of the lead, given in Eq.~(\ref{self-G-inf}).
Here, special attention must be paid to the inversion symmetry breaking induced by the interchain coupling terms.
This symmetry breaking plays a crucial role in our results and will be explicitly demonstrated in Section \ref{Sec:Non-eq_LR}.
Due to this symmetry breaking, the current contributions from both the left and right edges of the lead must be considered when deriving the differential conductance.
To account for this, we introduce the averaged current, defined as $I_{\beta_{\alpha}\beta'_{\alpha'}} = \frac{1}{2}\lb I_{\beta_{\alpha}} + I_{\beta'_{\alpha'}} \rb$, where the indices $\beta$ and $\beta'$ correspond to the normal ($\beta=\rm N$) or superconducting ($\beta = \rm S$) chains, and $\alpha$ and $\alpha'$ denote the left ($\alpha = \rm L$) or right ($\alpha = \rm R$) edge of the chains.
The current $I_{\beta_\alpha}:= \expval{\dot{N}_{\beta_\alpha}}$ is the current at $\alpha$ edge of the $\beta$ chain, where $N_{\beta_{\alpha}}$ is the electron number operator defined on the edge site.
The differential conductance from this average current is expressed as follows;
\begin{align}
    &\pdv{I_{\beta_{\alpha}\beta'_{\alpha'}}}{V} 
    \notag \\ 
    &= 
    \frac{e^2}{4h}\sum_{E=\pm V/2} \left[
     \varGamma_{\beta_{\alpha}}^{-} \lb
    \varGamma_{\beta'_{\alpha'}}^{-} \abs{\check{G}^{r}_{\beta_{\alpha}^{+},\beta_{\alpha'}^{\prime-}}}^2 
    + \varGamma_{\beta_{\alpha}}^{+}\abs{\check{G}^{r}_{\beta_{\alpha}^{+},\beta_{\alpha}^{-}}}^2
    \rb,
    \right.
    \notag \\ 
    &\quad +
    \left.
    \varGamma_{\beta'_{\alpha'}}^{-} \lb
    \varGamma_{\beta_{\alpha}}^{-} \abs{\check{G}^{r}_{\beta_{\alpha'}^{\prime+},\beta_{\alpha}^{-}}}^2 
    + \varGamma_{\beta'_{\alpha'}}^{+}\abs{\check{G}^{r}_{\beta_{\alpha'}^{\prime+},\beta_{\alpha'}^{\prime-}}}^2
    \rb
    \right],
    \label{eq_dI_dV}
\end{align}
where $\check{G}^{r}_{\beta_{\alpha}^{+},\beta_{\alpha'}^{\prime-}}$ is the component of the retarded Green's function, and its index $\pm$ represents the particle ($+$) or hole ($-$) part.
The $\varGamma_{\beta'_{\alpha'}}^{-}$ is the imaginary part of the edge self-energy.
These functions are defined as 
\begin{align}
    &\check{G} \equiv \LB z\check{I}_{4N} - \check{H}_{\rm{couple}} -\check{\Sigma}^{\rm \beta}_{\alpha} -\check{\Sigma}^{\rm \beta'}_{\alpha'} \RB^{-1},
    \notag \\
    &\check{\varGamma}_{\beta_\al} = -2\Im \check{\Sigma}_{\al}^{\beta,r},
    \\
    &\check{F}_{\beta_\al} = I_{\rm N} \otimes \mqty(
    f(\hbar\omega-eV_{\beta_\al}) & 0 \\
    0                   & f(\hbar\omega+eV_{\beta_\al})
    ),
    \\
    &\check{\Sigma}_{\al}^{\beta,<} = \check{\varGamma}_{\beta_\al} \check{F}_{\beta_\al},
    \\
    &\check{G}^{<} = \check{G}^r (\check{\Sigma}_{\alpha}^{\beta,<}+\check{\Sigma}_{\alpha'}^{\beta',<}) \check{G}^a.
\end{align}
The derivation and detailed formulation of the differential conductance are provided in Appendix \ref{sec_App_A}.

%% file: Numerical.tex
\begin{figure}
    \centering
    \includegraphics[scale=1.4]{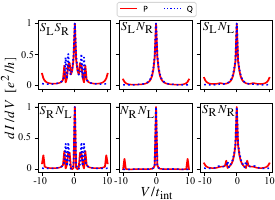}
    \caption{
    The differential conductance of all lead connecting patterns.
    Q and P denote the quasiperiodic and periodic, respectively.
    For the periodic case, we set $\rho_{\rm S}=\rho_{\rm N} = 1.0$.
    For the quasiperiodic case, we set $\rho_{\rm S} = 1/\tau_{\rm g}^2$ and $\rho_{\rm N} = 1.0$.
    Other parameters are set as
    $\Delta_{\rm int} = t_{\rm int} = 1.0$, $c_{\rm S}=10$, $c_{\rm N} = 0.8$, $\alpha = 0.3$, $\mu=0.0$ and $N=145$ (11th generation).
    }
    \label{V_cond_stack}
\end{figure}

In this section, we calculate the differential conductance of the coupled system shown in Fig.~\ref{Fig_stack}.
Before proceeding with the calculations, we make several assumptions regarding the model parameters.
First, we set the interchain hopping as the energy reference.
Second, we assume that the energy gap for Bogoliubov quasiparticles is determined by the average value of the real-space pair potential \cite{Kitaev_stack}.
Specifically, we impose the condition that the average pair potential, defined as $\alpha c_{\rm S/N} = \sum_{i=1}^{N}\Delta_{i,i+1}^{\rm S/N}/L_{N}$ remains constant, where $L_{N}$ is the total number of bond in $N$-site chain, and $\alpha \in [0,1]$ represents the ratio between the hopping and the pair potential corresponding to the distance from the critical point.
For the $n$th generation of Fibonacci quasicrystals, the bond number $L_{N}$ is represented by the $n$th Fibonacci number $F_n$ as $L_{N}=F_{n}$.
From the inflation rule, it follows that the $n$th generation of Fibonacci quasicrystals has $F_{n-1}$ long and $F_{n-2}$ short tiles.
Thus, the hopping amplitude and pair potential are given by
\begin{align}
     t_{\rm S/N} &= \frac{c_{\rm S/N} F_{n}}{F_{n-2}+\rho_{\rm S/N} F_{n-1}},
     \notag\\ 
     \Delta_{\rm S/N} &= \alpha t_{\rm S/N}, \quad 0 \leq \alpha \leq 1,
\end{align}
where $c_{\rm S}$ and $c_{\rm N}$ denote the average of the pair potential of the superconducting and normal chain, respectively.
The parameter $\rho_{\rm S/N}$ is the ratio of the hopping amplitude between the long and short tiles.

According to these assumptions, we present the numerical results for the differential conductance in the coupled system in Fig.~\ref{V_cond_stack}.
In each panel, the labels $S_{\rm L}$ and $S_{\rm R}$ denote that the lead is connected to the left and right edge of the superconducting chain, and $N_{\rm L}$ and $N_{\rm R}$ correspond to the left and right edge of the normal chain. 
The solid and dotted lines represent the cases of periodic and quasiperiodic chains, denoted as P and Q, respectively.
This figure demonstrates that the zero-bias conductance peak with the value of $e^2/h$ appears for all lead connection patterns shown here in both periodic and quasiperiodic chains.
Therefore, within the parameter regimes shown in Fig.~\ref{V_cond_stack}, quasiperiodicity does not significantly alter the zero-bias conductance peak associated with the Majorana edge mode.

%% file: Invariant.tex
In this section, we discuss the topological invariant for the coupled system.
Since the coupled Kitaev chain belongs to the BDI topological class, the topological index based on the pseudospectral method can be used to classify their topological phases \cite{K-theory_pseudo, saito2024, Fulga-topo-SC}.
First, we consider the Hamiltonian $H_{\rm couple}$ with open boundary conditions (OBC) defined in Eq.~(\ref{Ham_Stack_mat}).
Here, we define the Pauli matrices $\tau_\mu$ in the Nambu basis as 
\begin{equation}
    \widehat{\beta}_{\Delta}^{\rm S/N}
    = -t_{i,i+1}^{\rm S/N}\tau_{z} -i \Delta_{i,i+1}^{\rm S/N}\tau_{y},
\end{equation}
where the definition of $\widehat{\beta}_{\Delta}^{\rm S/N}$ is in Eq.~(\ref{Ham_Stack_mat}).
According to this definition, the chiral operator satisfying the relationship $H_{\rm couple}\hat{\varGamma} = -\hat{\varGamma} H_{\rm couple}$ can be given as
\begin{align}
    \hat{\varGamma}^{\rm S/N} &= \bigoplus_{i=1}^{N_{\rm S/N}}\tau_{x},
    \notag\\ 
    \hat{\varGamma} &= \hat{\varGamma}^{\rm S} \oplus \hat{\varGamma}^{\rm N},
\end{align}
where $N_{\rm S/N}$ is the number of superconducting or normal chain sites.

Using this operator, we can define the topological index as
\begin{align}
    \nu = \frac{1}{2}\mathrm{sig}\LB \lb X + iH_{\rm couple} \rb \hat{\varGamma} \RB \in \mathbb{Z},
\end{align}
where the "sig" is the signature of a matrix defined as the difference between the number of positive and negative eigenvalues.
The matrix $X$ is the position operator with the Nambu basis, i.e.
\begin{align}
    X^{\rm S/N} &= \mathrm{diag}(x_{1}^{\rm S/N},x_{1}^{\rm S/N}, \cdots x_{N_{\rm S/N}}^{\rm S/N},x_{N_{\rm S/N}}^{\rm S/N}),
    \notag\\
    X &= X^{\rm S} \oplus X^{\rm N}
\end{align}
where attention must be paid to the definition of coordinates.
This paper defines the coordinates as $x_{i}^{\rm S/N}\in [-0.5,0.5]$.

%% file: Phase.tex
Here, we discuss the properties of the topological phase transition in the coupled system.
The topological phase transition is characterized by the change of the topological invariant, which corresponds to the number of Majorana fermions at the edges of the system.
Now, we have four edges in the system. However, the number of Majorana fermions is not always four.
So, we used the local differential conductance to detect whether there is a Majorana fermion at the edge or not.
The local differential conductance is defined as
\begin{align}
    \pdv{I_{\beta_{\alpha}}}{V}
    = \frac{e^2}{h}\Gamma^{-}_{L}
    \lb \Gamma^{-}_{R}|\check{G}^{r}_{\beta_{\alpha},\beta_{\alpha}}|^2 + \Gamma^{+}_{L}|\check{G}^{r}_{\beta_{\alpha},\beta_{\alpha}+1}|^2 \rb,
\end{align}
where $\beta_{\alpha}$ is the site index of edges.

\subsection{Phase transition along with chemical potentials}
\begin{figure}[!htbp]
    \centering
    \includegraphics{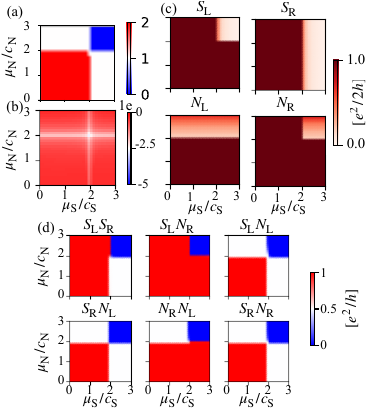}
    \caption{Topological phase transition of the coupled system with periodic chains.
    The parameters are set as $\rho_{\rm S} = \rho_{\rm N} = 1.0$, $c_{\rm S} = 10.0$, $c_{\rm N} = 1.0$ and $\alpha = 0.3$.
    (a) is the topological invariant of this system. The red, white, and blue regions correspond to the topological phases with $\nu = 2$, $1$, and $0$, respectively.
    (b) The energy spectrum with periodic boundary conditions is shown with a logarithmic scale.
    (c) is the local differential conductance of the system renormalized by $e^2/2h$.
    The labels $S_{\rm L}$, $S_{\rm R}$, $N_{\rm L}$, and $N_{\rm R}$ correspond to the left edge of the superconducting chain, the right edge of the superconducting chain, the left edge of the normal chain, and the right edge of the normal chain, respectively.
    (d) is the nonlocal differential conductance of the system renormalized by $e^2/h$.
    }
    \label{Fig:phase-PP}
\end{figure}

This section considers the correspondence between the topological phase transition and the chemical potential.
Here, the chemical potential of superconducting and normal chains is changed independently.
In Fig.~\ref{Fig:phase-PP}, we show the topological phase transition of the coupled system with periodic chains.
In panel (a), we can observe the topological phases with $\nu = 2$, $1$, and $0$.
At the phase boundary, the energy spectrum with periodic boundary conditions closes the gap, which is shown in panel (b).
Panels (c) and (d) show the local and nonlocal differential conductance of the system, respectively.
In the region with $\nu=2$, the local and nonlocal differential conductance have values of $e^2/2h$ and $e^2/h$, respectively.
Therefore, all four edges have Majorana fermions.
Conversely, in the region with $\nu=0$, there is no Majorana fermion in the system because the local and nonlocal differential conductance have values of $0$.
On the other hand, there are two types of topological phases with $\nu=1$.
In the region with $\nu=1$ and $\mu_{\rm N}/c_{\rm N} > 2$, the local differential conductance at $S_{\rm L}$, $S_{\rm R}$, and $N_{\rm R}$ have values of $e^2/2h$.
Additionally, the nonlocal differential conductance has a value of $e^2/h$ with patterns $S_{\rm L}S_{\rm R}$ and $S_{\rm L}N_{\rm R}$, while has a value of $e^2/2h$ with other patterns.
Therefore, there is a Majorana fermion at the left edge of the superconducting chain, and the Majorana fermion at the right edge is distributed between the superconducting and normal chains.
Considering that this region is the non-trivial phase of the superconducting chain, we can conclude that the wave function of the Majorana fermion at the right edge seeps into the normal chain from the superconducting chain.
In comparison, the phase with $\mu_{\rm S}/c_{\rm S} > 2$ hosts the local differential conductance with value of $e^2/2h$ at the edges $N_{\rm R}$, $N_{\rm L}$ and $S_{\rm L}$.
Furthermore, nonlocal conductance has a value of $e^2/h$ with the patterns $S_{\rm L}N_{\rm R}$ and $N_{\rm R}N_{\rm L}$.
By parallel discussion, therefore, there is an isolated Majorana fermion at the right edge of the normal chain, and there is the seepage of Majorana fermion's wave function into the left edge of the superconducting chain from the left edge of the normal chain.
These two $\nu=1$ regions have the same topological index but different Majorana distributions.
In one region, one Majorana mode is localized at the left edge of the superconducting chain, whereas the right-edge Majorana mode is distributed over the superconducting and normal chains.
In the other region, the localization and seepage pattern is shifted to the opposite chain/edge combination.
Therefore, the integer invariant $\nu$ specifies the number of Majorana pairs but does not uniquely determine their spatial distribution in the coupled system.
The inequivalence of the left and right edges is explained by the inversion breaking discussed in the section \ref{Sec:Non-eq_LR}.

\begin{figure}[!htbp]
    \centering
    \includegraphics{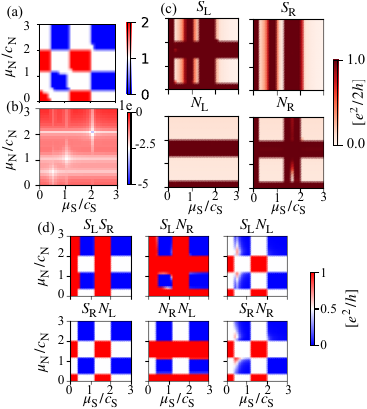}
    \caption{Topological phase transition of the coupled system with quasiperiodic chains.
    The parameters are set as $\rho_{\rm S} = \rho_{\rm N} = 1/\tau_{\rm g}^2$ and $\alpha = 0.3$.
    What is shown in each panel is the same as in Fig.~\ref{Fig:phase-PP}.
    }
    \label{Fig:phase-QQ}
\end{figure}

\begin{figure}[!htbp]
    \centering
    \includegraphics{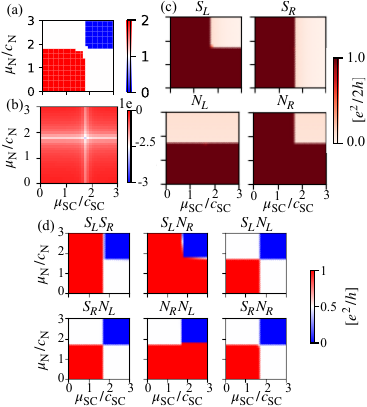}
    \caption{Topological phase transition of the coupled system with quasiperiodic chains with $\alpha = 1.0$.
    Other parameters are the same as in Fig.~\ref{Fig:phase-QQ}.
    What is shown in each panel is the same as in Fig.~\ref{Fig:phase-PP}.
    }
    \label{Fig:phase-QQ-alpha1}
\end{figure}

Next, we consider the quasiperiodic system in Fig.~\ref{Fig:phase-QQ}.
We can see the effect of quasiperiodicity by comparing it with Fig.~\ref{Fig:phase-PP}.
In the regions of the phase of $\nu = 2$ for the periodic case, we can find the new topological phase transitions and the phases with $\nu = 1$ or $\nu = 0$.
In addition, in the regions with $\nu = 1$ for the periodic case, the trivial phase with $\nu = 0$ appears, and topological phase transitions occur.
Although these new properties exist in a phase transition, the properties of the phase with $\nu = 1$, such as isolation or seepage of the Majorana fermion, are the same as those of a coupled system with periodic chains.
Furthermore, we show the calculations with $\alpha = 1.0$, corresponding to the short coherence length limit, in Fig.~\ref{Fig:phase-QQ-alpha1}.
According to this figure, the new topological phase transition disappears.
Therefore, quasiperiodicity affects the topological phase when the coherence length is sufficiently long.

\begin{figure}[!htbp]
    \centering
    \includegraphics{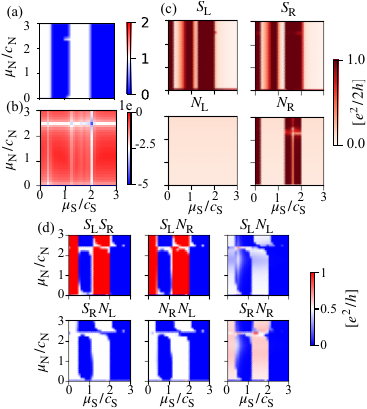}
    \caption{Topological phase transition of the coupled system with quasiperiodic chains with $\rho_{\rm S} = 1/\tau_{\rm g}^2$, $\rho_{\rm N} = \rho_{\rm S}^{3.99}$ and $\alpha = 0.3$.
    Other parameters are the same as in Fig.~\ref{Fig:phase-QQ}.
    What is shown in each panel is the same as in Fig.~\ref{Fig:phase-PP}.
    }
    \label{Fig:phase-QQp}
\end{figure}

Next, we discuss the quasiperiodic case with $\rho_{\rm N} = \rho_{\rm S}^{3.99}$ presented in Fig.~\ref{Fig:phase-QQp}, which is the most robust case discussed in Ref.~\cite{Kitaev_stack}.
According to panel (a), we cannot find the topological phase with $\nu = 2$ and can only see the phase with $\nu = 1$ and the trivial phase $\nu = 0$.
About the topological phase with $\nu = 1$, only the phase that hosts an isolated Majorana fermion at the left edge of the superconducting chain and distributes one at the right-hand side appears.

\clearpage

\subsection{Inequivalence of left and right edges}\label{Sec:Non-eq_LR}

In our results, it is non-trivial that while the Majorana fermion is on the right edge of the normal Kitaev chain, there is no Majorana fermion on the left edge.
Intuitively, one would expect Majorana modes on both edges in the P-P case, but their presence is asymmetric even in this case.
To explain this behavior, we consider the inversion symmetry $\hat{I}$ defined as
\begin{align}
    \hat{I}d^{\rm SC/N}_{j}\hat{I}^{-1} = id^{\rm SC/N}_{-j},
    \quad 
    \hat{I}d^{\rm SC/N \dag}_{j}\hat{I}^{-1} = -id^{\rm SC/N \dag}_{-j},
\end{align}
where the coefficient $i$ is the imaginary unit.
According to this definition, the Hamiltonian corresponding to one Kitaev chain in the coupled system defined in Eq.~(\ref{Ham_stack_SN}) satisfies that
\begin{align}
    \hat{I}H_{S/N}\hat{I}^{-1} = H_{S/N}.
\end{align}
Therefore, this Hamiltonian is inversion symmetric, i.e., the single periodic Kitaev chain is.
However, the interchain terms $H_{int}$ defined by Eq.~(\ref{Ham_stack_int}) satisfy that
\begin{align}
    \hat{I}H_{\rm int}\hat{I}^{-1}
    =\sum_{i=1}^{N}\LB
    -\tau_{\rm int} d_{i}^{S\dag}d_{i}^{N} 
    - \delta_{\rm int} d_{i}^{S}d_{i}^{N} + h.c.
    \RB.
\end{align}
Thus, this Hamiltonian breaks the inversion symmetry.
Therefore, the unbalance of the existence of Majorana fermion is justified by this inversion breaking.
This inversion breaking cannot be removed by a redefinition of the $U(1)$ phase or by shifting the inversion center, because the interchain hopping and interchain pairing terms transform with incompatible relative signs.
For example, if we use the inversion defined as
\begin{align}
    \hat{I}c^{SC}_{j}\hat{I}^{-1} \to ic^{SC}_{-j},
    \quad
    \hat{I}c^{N}_{j}\hat{I}^{-1} \to -ic^{N}_{-j},
\end{align}
then the terms of $\delta_{\rm int}$ are symmetric, but the terms of $\tau_{\rm int}$ are not symmetric.
Therefore, we cannot define the inversion symmetry with the interchain hopping and pair potential.
This inversion breaking justifies that the asymmetric seepage of the Majorana mode's wave function exists even in the periodic system.

%% file: equivalence.tex
This section proves that the quasiperiodic Kitaev chain is also equivalent to an $s$-wave proximity-induced system with strong spin-orbit coupling (SOC).
The method to derive the effective model of the Kitaev chain is provided by Wang {\it et al}.~\cite{Wang_eff_model}.
This method can derive the effective $k\cdot p$ Hamiltonian around the $\Gamma$ point of the pseudo-Brillouin zone to map from higher dimensional space called hyperspace to physical space.
The maps considered in this method must have the same rank value as the dimensions of hyperspace.
In the case of Fibonacci quasicrystal, the dimensions of hyperspace are two, and the angle of the map is the value of an irrational number \cite{Fahannathan_2021}.

\subsection{Effective Hamiltonian}

First of all, we derive the effective Hamiltonian of Fibonacci quasicrystal with SOC around $\Gamma$ point of the pseudo-Brillouin zone.
If we assume the existence of Rashba-type SOC, then the Hamiltonian in one-dimensional (1D) physical space is written as 
\begin{equation}
    \begin{split}
        H=& \sum_{i}\e c^{\dag}_{i}c_{i} 
          + \sum_{<i,j>}t(\vv{d}_{ij})c^{\dag}_{i}c_{j} \\
          &+ i\sum_{<i,j>}\la(\vv{d}_{ij})(c^{\dag}_{i}\sigma_{y}c_{j}-c^{\dag}_{j}\sigma_{y}c_{i}),
    \end{split}
\end{equation}
where $c^{\dag}_{i}=(c^{\dag}_{i\al\up},c^{\dag}_{i\al\down})$ are electron creation operator at the $i$ th site. In the first term, $\e$ is on-site energy. The second term is the hopping term with hopping parameter $t(\vv{d}_{ij})$, which depends on vector $\vv{d}_{ij}$ between site $i$ and $j$.
$\la$ is constant and represents SOC strength, and $s_y$ is $y$ component of the Pauli matrix operating to spin degrees of freedom.
In the second and third terms, sums over the nearest-neighbor sites are taken.

We perform novel Fourier expansion to obtain the effective Hamiltonian of Fibonacci quasicrystal in hyperspace.
The Fibonacci quasicrystal is generated by projecting a square lattice in two-dimensional hyperspace to 1D physical space using projection operator $S=(1,1/\tau_{\rm g})$, where $\tau_{\rm g}$ is the golden ratio $(1+\sqrt{5})/2$.
We introduce the wave vector in hyperspace as $\vv{\Pi}=(k_{x},k_{y})^T$, and we can write the Fourier expansion as
\begin{align}
    c^{\dag}_{i} &= \sum_{\vv{\Pi}}c^{\dag}_{\vv{\Pi}} e^{i\LB (S\vv{\Pi})^{T}\cdot\vv{r}_i \RB}, \\
    c_{i} &= \sum_{\vv{\Pi}}c_{\vv{\Pi}} e^{-i\LB (S\vv{\Pi})^{T}\cdot\vv{r}_i \RB}.
\end{align}
Using this expansion, each terms of Hamiltonian are written as
\begin{align}
    &\sum_{i}\e c^{\dag}_{i}c_{i} = \sum_{\vv{\Pi}}\e c^{\dag}_{\vv{\Pi}}c_{\vv{\Pi}}, \\
    &\sum_{<i,j>}t(\vv{d}_{ij})c^{\dag}_{i}c_{j}
    = 2t\LB P_{0}\cos(k) + P_{1}\tau_{\rm g}^2 \cos\lb \frac{k}{\tau_{\rm g}} \rb \RB c^{\dag}_{\vv{\Pi}}c_{\vv{\Pi}}
\end{align}
\begin{equation}
    \begin{split}
      &i\sum_{<i,j>}\la(\vv{d}_{ij})(c^{\dag}_{i}\sigma_{y}c_{j}-c^{\dag}_{j}\sigma_{y}c_{i}) \\
      &= 2\la \LB P_{0}\sin(k) + P_{1}\tau_{\rm g}^2 \sin(k/\tau_{\rm g}) \RB c^{\dag}_{\vv{\Pi}}\sigma_{y}c_{\vv{\Pi}},
    \end{split}
\end{equation}
where $P_0$ and $P_1$ are the statistical average distribution of inter-atomic vectors in quasiparticles corresponding to long distance and short distance, respectively.
They are called the Patterson function and can be determined by diffraction data.
In above calculations, we assume that hopping parameter $t(\vv{d_{ij}})$ and SOC parameter $\la(\vv{d}_{ij})$ are proportional to $1/\tau_{g}^2$. 
Now, we can get the effective Hamiltonian around $\Gamma$ point of pseudo-Brillouin zone as
\begin{align}
    H_{\rm{eff}}(k) &= \e_{0} + a k^2 + b\si_{y}k, \\
    \e_{0} &= \e + 2t(P_{0}+P_{1}\tau_{\rm g}^2), \\
    a &= -t(P_{0}+P_{1}), \\
    b &= -2\la(P_{0}+P_{1}\tau_{\rm g}).
\end{align}
Additionally, if the Zeeman field is added, then the effective Hamiltonian is written as
\begin{align}
    H_{\rm{eff}}(k) &= \e_{0} + a k^2 + b\si_{y}k + M_{z}\sigma_{z},
\end{align}
where $M_z$ is the strength of the Zeeman field parallel to the $z$-axis.
The above Hamiltonian is equivalent to the Hamiltonian of a one-dimensional semiconductor with Rashba-type SOC with corrections of parameters due to the quasiperiodicity.

\subsection{Proximity effect of \texorpdfstring{$s$-wave}{s-wave} superconductivity}
This section clarifies the equivalence between our model in the proximity effect and the quasiperiodic Kitaev chain.
First, we consider the Nambu formula of our effective Hamiltonian with the $s$-wave proximity effect. 
This Hamiltonian is written as 
\begin{align}
    H^{\rm MF}(k) =
    \mqty(  H_{\rm eff}(k)      & i\Delta_{0}\si_{y} \\
        -i\Delta^{*}_{0}\si_{y} &   -H_{\rm eff}^{*}(-k) 
    ).
\end{align}
Now, we write the eigenvalues of effective Hamiltonian as
\begin{align}
    \xi_{\pm} = \e_{0} + ak^2 \pm \sqrt{b^2k^2+M_{z}^2} 
\end{align}
Using this, we can write the eigenvalues of Hamiltonian $H^{\rm MF}$ as
\begin{align}
    E^2 = \frac{\xi_{+}^2 + \xi_{-}^2}{2} + \Delta_{0}^2 \pm \sqrt{\frac{(\xi_{+}^2 - \xi_{-}^2)^2}{4} +4M_{z}^2\Delta_{0}^2 }.
\end{align}
At $k=0$, the bulk gap closes when
\begin{align}
    M_{z}^2 = \e_{0}^2+\Delta_{0}^2.
\end{align}
Thus, within the low-energy Rashba-wire description, the topological phase is realized for
\begin{align}
    |M_z| > \sqrt{\e_{0}^2+\Delta_{0}^2}.
\end{align}

Next, we show the equivalence between this and the Kitaev chain models.
To do this, we transform the basis of $H^{\rm MF}$ to the basis that diagonalize $H_{\rm eff}(k)$.
Therefore, the Hamiltonian is rewritten as
\begin{align}
    H^{\p}(k) = 
    \mqty(
    \xi_{+}             &       0           &  -i\Delta_{\rm odd}(k)    &   -\Delta_{\rm even} \\
        0               &   \xi_{-}         &    \Delta_{\rm even}      &    i\Delta_{\rm odd}(k) \\
    i\Delta_{\rm odd}(k)    &\Delta_{\rm even}      &       -\xi_{+}        &       0       \\
    -\Delta_{\rm even}      &-i\Delta_{\rm odd}(k)  &           0           &       -\xi_{-}
    ),
\end{align}
where $\Delta_{\rm even}$ and $\Delta_{\rm odd}(k)$ are defined as
\begin{align}
    \Delta_{\rm even}
    &=
    \frac{M_{z}\Delta_{0}}{\sqrt{b^2k^2+M_{z}^2}},
    \notag\\
    \Delta_{\rm odd}(k)
    &=
    \frac{b\Delta_{0}k}{\sqrt{b^2k^2+M_{z}^2}},
\end{align}
respectively.
Now, considering the spin splitting due to the Zeeman field, only one spin sector contributes to the superconductivity.
Therefore, we can reduce the Hamiltonian to $2\times 2$ matrices as
\begin{align}
    H_{\rm eff}^{\p}(k) =
    \mqty(
        \xi_{+}         & -i\Delta_{\rm odd}(k) \\
    i\Delta_{\rm odd}(k)    &       -\xi_{+}
    ).
\end{align}
This is just the Bogoliubov Hamiltonian of spinless $p$-wave superconductivity.
Therefore, our model of a quasiperiodic Kitaev chain can be realized by an $s$-wave-proximity-induced quasiperiodic chain with SOC and a Zeeman field.

%% file: Appendix/Single-chain.tex
In this section, we derive the formula of differential conductance of a single Kitaev chain and the system coupling two Kitaev chains.
For the coupled system, we consider all patterns to connect leads to arbitrary two edges of coupled Kitaev chains.
The effect of connecting with lead is taken into real-space Green's function defined by tight-binding Hamiltonian of finite size Kitaev chain with open boundary condition as local self-energy at the edge calculated by recursive Green's function of half-infinite leads \cite{Linear_nonLinear_Kitaev,Self_ene_Recursive}. 
In the following, the $2\times 2$ matrix is denoted by $\widehat{A}$, and the $2N\times 2N$ matrix is denoted by $\widetilde{A}$.

\begin{figure}
    \centering
    \includegraphics[scale=0.5]{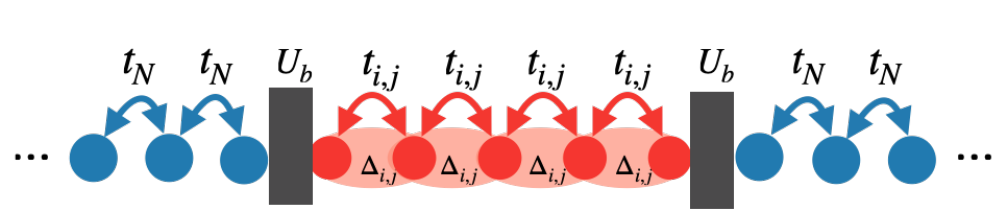}
    \caption{Schematic drawing of the case that connects a single Kitaev chain and leads.
    In this figure, the arrows and ellipses represent the hopping and coupling between nearest-neighbor sites, respectively.
    The black rectangles are the potential barrier between the leads and the Kitaev chain.
    If this system has quasiperiodicity, the nearest-neighbor hopping and coupling depend on the distance between the neighbor sites.
    }
    \label{Fig_single}
\end{figure}

\subsection{Model}

We consider the N-S-N system constructed by half-infinite normal metal lead and finite quasiperiodic Kitaev chain illustrated in Fig.~\ref{Fig_single}.
The tight-binding Hamiltonian of this system is read as follows;
\begin{equation}
    \begin{split}
        H_{\al} =& \sum_{i=0}^{\pm\infty}\LB \mu c_{\al,i}^\dag c_{\al,i} - t_{\rm N} \lb c_{\al,i+1}^\dag c_{\al,i} + h.c. \rb \RB,
    \end{split}
\end{equation}
\begin{equation}
    \begin{split}
        H_{\rm S} =& \sum_{i=1}^{N}\LB \mu d_{i}^\dag d_{i}\RB \\
        &+ 
        \sum_{i=1}^{N-1}\LB -t_{i,i+1} d_{i+1}^\dag d_{i} + \Delta_{i,i+1}d_{i}d_{i+1}  + h.c. \RB,
        \label{dif_Kitaev_single}
    \end{split}
\end{equation}
\begin{equation}
    \begin{split}
        H_{T} =& \sum_{\al=R,L}t_{\al}(c_{\al,0}^{\dag}d_{1} + d_{1}^\dag c_{\al,0} ),
    \end{split}
\end{equation}
where $H_{\alpha}$ $(\alpha=L, R)$, $H_{\rm S}$ and $H_T$ are the Hamiltonian of half-infinite normal metal leads connected into left and right edge, the Hamiltonian of superconducting Kitaev chain with $N$ sites, and the tunneling Hamiltonian between Kitaev chain and leads at both edges of Kitaev chain, respectively.
The operator $c_{\alpha,i}$ ($c_{\alpha,i}^\dag$) is the annihilation (creation) operator of the electron in the normal metal leads defined on the $i$th site, and
$d_{\alpha,i}$ ($d_{\alpha,i}^{\dag}$) is the annihilation (creation) operator of the electron in the Kitaev chain.
If the Kitaev chain has quasiperiodicity, then the values of hopping parameter $t_{i,i+1}$ and pair potential $\Delta_{i,i+1}$ depend on the distance between the nearest-neighbor sites.
For convenience, we write the above Hamiltonians on the particle-hole basis as follows
\begin{equation}
    \begin{split}
        H_{\al} =& \mqty(c_{\al}^\dag & c_{\al} ) \widetilde{H}_{\al} \mqty(c_{\al} \\ c_{\al}^\dag ),
        \\
        \widetilde{H}_{\al} =& \mqty(
        \widehat{\mu}              & \widehat{\beta}_{\al} &                        & \\
        \widehat{\beta}_{\al}^\dag & \ddots            & \ddots                 & \\
                               & \ddots            & \ddots                 & \widehat{\beta}_{\al} \\
                               &                   & \widehat{\beta}_{\al}^\dag & \widehat{\mu}  
        ),
        \\
        \widehat{\mu} =& \mqty(
        -\mu & 0 \\
        0    & \mu
        ),\quad
        \widehat{\beta}_{\al} = \mqty(
        -t_{\rm N} & 0 \\
        0    & t_{\rm N}
        ),
    \end{split}
\end{equation}
\begin{equation}
    \begin{split}
        H_{\rm S} =& \mqty(d^\dag & d ) \widetilde{H}_{\rm S} \mqty(d \\ d^\dag ),
        \\
        \widetilde{H}_{\rm S} =& \mqty(
        \widehat{\mu}              & \widehat{\beta}_{\Delta} &                        & \\
        \widehat{\beta}_{\Delta}^\dag & \ddots            & \ddots                 & \\
                               & \ddots            & \ddots                 & \widehat{\beta}_{\Delta} \\
                               &                   & \widehat{\beta}_{\Delta}^\dag & \widehat{\mu}  
        ),
        \\
        \widehat{\beta}_{\Delta} =& \mqty(
        -t_{i,i+1}        & -\Delta_{i,i+1} \\
        \Delta_{i,i+1}    & t_{i,i+1}
        ).
    \end{split}
\end{equation}

\subsection{Formulations of differential conductance}
In this section, we derive the real space of Green's of the finite-size quasiperiodic Kitaev chain with leads at both edges.
First of all, we derive the surface Green's function at the end of left and right leads by the recursive Green's function method \cite{Takagi_Tamura_2020,Asano_Tanaka_2013}.
This method is so useful for the derivation of surface Green's function of a half-infinite system.
Now, we consider the  $2N \times 2N$ Green's function defined as
\begin{align}
    G_{\al}^{(N)} = \lb z\widetilde{I}_{2N} - \widetilde{H}_{\al} \rb^{-1},
\end{align}
where $\widetilde{I}_{2N}$ is the $2N \times 2N$ identity matrix.
Using the M\"{o}bius transform defined as
\begin{align}
    \mqty(
    A & B \\
    C & D
    ) \bullet Y \equiv (AY+B)(CY+D)^{-1} ,
\end{align}
the $(N,N)$ component of Green's function $G_{\al}^{(N)}$ as $2\times2$ matrix is written as
\begin{align}
    (G_{\al}^{(N)})_{N,N} &= X_\al \bullet (G_{\al}^{(N-1)})_{N-1,N-1} \\
    X_\al &= \mqty(
    \widehat{O}_2                  & \widehat{\beta}_{\al}^{-1} \\
    -\widehat{\beta}_{\al}^{\dag}  & (z\widetilde{I}_2 -\widehat{\mu})\widehat{\beta}_{\al}^{-1}
    ),
\end{align}
where $\widehat{O}_2$ is the $2\times2$ zero matrix.
The complex value $z$ becomes $z = E + i\delta$ for the retarded Green's function, and $z = E - i\delta$ for the advanced Green's function, where $E$ is the energy of the electron and $\delta$ is the infinitesimal positive value.
To repeat this transformation, we get
\begin{align}
    (G_{\al}^{(N)})_{N,N} &= X^{N-1} \bullet (G_{\al}^{(1)})_{1,1},
    \\
    (G_{\al}^{(1)})_{1,1} &= \lb z\widehat{I}_2 - \widehat{\mu} \rb^{-1},
\end{align}
where $\widehat{I}$ is the $2\times2$ identity matrix.

If we define the eigenvalues and eigenvectors as
\begin{align}
    (Q_\al)^{-1}X_{\al}Q_\al
    &= \rm{diag} (\la_1, \la_2,\la_3,\la_4),
\end{align}
where $\la_1 \leq \la_2 < \la_3 \leq \la_4$,
then the surface Green's of the half-infinite system is derived by considering the limit $N\to \infty$ as
\begin{align}
    \widehat{G}_{\al}^{\infty} = (\widehat{Q}_{\al})_{12}(\widehat{Q}_\al)_{22}^{-1}.
\end{align}
Using Green's function, we can define the self-energy by contacting with lead as discussed in Ref.~\cite{Self_ene_Recursive} 
\begin{align}
    \widehat{\Sigma}_{\al} = \widehat{\beta}_{\al}^\dag \widehat{G}_{\al}^{\infty} \widehat{\beta}_\al. 
    \label{self-G-inf}
\end{align}

Following the formulations in \cite{Linear_nonLinear_Kitaev}, we define the self-energy of the finite Kitaev chain as
\begin{align}
    \widetilde{\Sigma}_{L}
    &= \mqty[
    \widehat{\Sigma}_{L} & \widehat{O}_2  & \cdots & \widehat{O}_2 \\
    \widehat{O}_2  & \widehat{O}_2  & \cdots & \widehat{O}_2 \\
    \vdots     & \vdots     & \ddots & \vdots \\
    \widehat{O}_2  & \widehat{O}_2  & \cdots & \widehat{O}_2 
    ], 
    \\
    \widetilde{\Sigma}_{R}
    &= \mqty[
    \widehat{O}_2  & \cdots & \widehat{O}_2  & \widehat{O}_2 \\
    \vdots     & \ddots & \vdots     & \vdots \\
    \widehat{O}_2  & \cdots & \widehat{O}_2  & \widehat{O}_2 \\
    \widehat{O}_2  & \cdots & \widehat{O}_2  & \widehat{\Sigma}_{R}
    ],\label{self-ene-single-2N}
\end{align}
where $\widehat{O}_n$ is the $n\times n$ zero matrix.
Finally, we consider the potential barrier between the lead and Kitaev chain.
This potential barrier works to shift the energy of electrons at the end of the Kitaev chain.
In other words, this potential is added to the real part of self-energy due to the contacting lead and Kitaev chain.
Now, we reach the Green's function of a quasiperiodic Kitaev chain with finite length.
The Green's function is written as
\begin{align}
    \widetilde{G} &\equiv \LB z\widetilde{I}_{2N} - \widetilde{H}_{\rm S} - \widetilde{\Sigma}_{L} - \widetilde{\Sigma}_{R} - \widetilde{U}_b \RB^{-1}, 
    \notag \\
    \widetilde{U}_b
    &=\mqty[
    \widehat{U}_b  & \cdots & \widehat{O}_2  & \widehat{O}_2 \\
    \vdots     & \ddots & \vdots     & \vdots \\
    \widehat{O}_2  & \cdots & \widehat{O}_2  & \widehat{O}_2 \\
    \widehat{O}_2  & \cdots & \widehat{O}_2  & \widehat{U}_{b}
    ],
    \notag \\
    \widehat{U}_{\rm b}
    &= \mqty(
    U_{\rm b} & 0 \\
    0   & -U_{\rm b}
    ),
    \label{def_G_2N2N}
\end{align}
where $U_{\rm b}$ is the potential barrier between the Kitaev chain and normal metal lead.

In this section, we derive the differential conductance in the formulation of Keldysh Green's function.
First of all, the current at the left edge is defined as $I_L = -e\ev{\dot{N}_L}$.
According to tunneling Hamiltonian, this current is written as
\begin{align}
    I_L = -\frac{ie}{\hbar}t_L \LB \ev{d_{1}^\dag c_{L,0}} - \ev{c_{L,0}^\dag d_{1}}  \RB.
\end{align}
This correlation function can be calculated by perturbation expansion of tunneling Hamiltonian.
We can write down as
\begin{align}
    I_L &= \frac{ie}{2}\int \frac{d\omega}{2\pi}
    \Tr \LB (I_{\rm N}\otimes\tau_z) \widetilde{\varGamma}_{L}\lB \widetilde{G}^{<} + \widetilde{F}_L(\widetilde{G}^r - \widetilde{G}^a) \rB \RB,
    \label{current_Keldysh}
    \\
    \widetilde{\varGamma}_{\al} &= -2\Im \widetilde{\Sigma}_{\al}^r,
    \\
    \widetilde{F}_\al &= I_{\rm N} \otimes \mqty(
    f(\hbar\omega-eV_\al) & 0 \\
    0                   & f(\hbar\omega+eV_\al)
    ),
    \\
    \widetilde{\Sigma}_{\al}^{<} &= \widetilde{\varGamma}_{\al} \widetilde{F}_\al,
    \\
    \widetilde{G}^{<} &= \widetilde{G}^r (\widetilde{\Sigma}_{L}^{<}+\widetilde{\Sigma}_{R}^{<}) \widetilde{G}^a,
\end{align}
where $I_N$ is the $N\times N$ identity matrix, and $f(\epsilon)$ is the Fermi distribution function.
In these definitions, the $\widetilde{G}^{<}$ and $\widetilde{\Sigma}^{<}$ are denote the lesser part of Green's function and self-energy, respectively.
The $\widetilde{G}^{r}$ and $\widetilde{G}^{a}$ are the retarded and advanced Green's function, respectively.
Samely, the $\widetilde{\Sigma}^{r}$ is the retarded part of the self-energy.
Next, our purpose is to simplify the current formula (\ref{current_Keldysh}).
In follows, we denote the $2N \times 2N$ Green's function as
\begin{align}
    \widetilde{G}
    =\mqty[
    \widehat{G}_{11} & \cdots & \widehat{G}_{1N} \\
          \vdots     & \ddots &     \vdots       \\
    \widehat{G}_{N1} & \cdots & \widehat{G}_{NN}
    ]_{2N \times 2N} .
\end{align}
In the first term of Eq.~(\ref{current_Keldysh}), only the $\widehat{G}^{<}_{11}$ need to be considered due to the trace in integral.
From the definition of lesser part $\widetilde{G}^{<}$, we can find as
\begin{align}
    \widehat{G}^{<}_{11} = \widehat{G}^{r}_{11}\widehat{\Sigma}_{L}\widehat{G}^{a}_{11} + \widehat{G}^{r}_{1N}\widehat{\Sigma}_{R}\widehat{G}^{a}_{N1} .
\end{align}
Additionally, in the second term of Eq.~(\ref{current_Keldysh}), we point out the equation
\begin{align}
    \widetilde{G}^{r} - \widetilde{G}^{a} = 
    -i\widetilde{G}^{r}\lb \widetilde{\varGamma}_{L} + \widetilde{\varGamma}_{R} \rb \widetilde{G}^{a},
\end{align}
where this equation comes from the definition (\ref{def_G_2N2N}).
According to these equations, the current is written as
\begin{align}
    I_{L}=& e\int \frac{d\omega}{2\pi}
    \left[
    \varGamma_{L}^{-}(\omega)\varGamma_{L}^{+}(\omega)\abs{\widetilde{G}^{r}_{1,2}(\omega)}^2 \lB f_{L}^{-}-f_{L}^{+} \rB
    \right. 
    \notag \\
    &+ \varGamma_{L}^{-}(\omega)\varGamma_{R}^{-}(\omega)\abs{\widetilde{G}^{r}_{1,2N-1}(\omega)}^2 \lB f_{L}^{-}-f_{R}^{-} \rB
    \notag \\
    &+\left. \varGamma_{L}^{-}(\omega)\varGamma_{R}^{+}(\omega)\abs{\widetilde{G}^{r}_{1,2N}(\omega)}^2 \lB f_{L}^{-}-f_{R}^{+} \rB
    \right],
\end{align}
where $f^{\pm}_{\alpha} = f(\hbar \omega \pm eV_{\alpha})$, and $\Gamma^+_{\alpha}(\omega)$ and $\Gamma^-_{\alpha}(\omega)$ denote the particle and hole components, respectively.
Using these equations, we can write the differential conductance with $V_L = -V_R = V/2$ as
\begin{align}
    \pdv{I}{V} = 
    \frac{e^2}{2h}\sum_{E=\pm V/2} \varGamma_{L}^{-} \lb
     \varGamma_{R}^{-} \abs{\widetilde{G}^{r}_{1,2N-1}}^2 + \varGamma_{L}^{+} \abs{\widetilde{G}^{r}_{1,2}}^2
    \rb, \label{Cond_single}
\end{align}
Once we reach this formulation, we can calculate numerically the differential conductance without calculating the whole Green's function \cite{nagai2017reduced}.

In this section, we briefly introduce this method.
One component of Green's function $\widetilde{G}_{i,j}$ can be represented as
\begin{align}
    \widetilde{G}_{i,j}
    = \vv{e}_{i}^{T} \widetilde{G} \vv{e}_{j},
\end{align}
where $\vv{e}_{i}$ is a vector such that
\begin{align}
    (\vv{e}_{i})_{\alpha}
    =\begin{cases}
    0 & \hbox{for } i \neq \alpha,
    \\
    1 & \hbox{for } i = \alpha,
    \end{cases}
\end{align}
and $\vv{e}_{i}^{T}$ is the transpose of vector $\vv{e}_{i}$.
Now, we define new vector $\vv{A}$ as
\begin{align}
    \vv{A} = \widetilde{G} \vv{e}_{j}.
\end{align}
In this definition, when we product the inverse matrix $\widetilde{G}^{-1}$ from left, it can be represented that
\begin{align}
    \widetilde{G}^{-1}\vv{A} =  \vv{e}_{j}.
\end{align}
Since $\widetilde{G}^{-1}$ and $\vv{e}_{j}$ are already well known, this equation is understood as the simultaneous equations of which the solution is $\vv{A}$.
Therefore, our purpose is to calculate Green's function as an inversion matrix in Eq.~(\ref{def_G_2N2N}), which comes down to the simple problem of solving simultaneous equations.

\subsection{Numerical results}
\begin{figure}
     \centering
     \subfigure[Differential conductance of periodic Kitaev chain]{
     \includegraphics[scale=0.75]{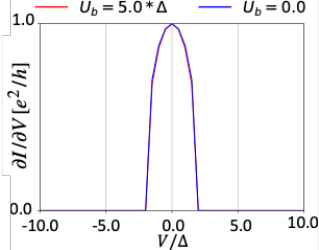}\label{Nume_single2_a}}
     \subfigure[quasiperiodic case]{
     \includegraphics[scale=0.75]{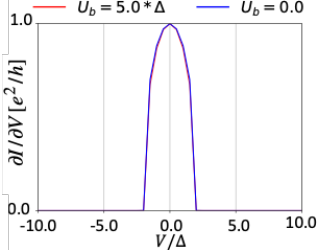}\label{Nume_single2_b}}
     \caption{Differential conductance for single Kitaev chain.
     These figures are calculated with parameters $c=10$, $t_{\rm N}=1.0$, $\mu=0.0$ and $N=2584$ (17th generation of Fibonacci quasicrystal).}
\end{figure}

First, we calculate the differential conductance of a single periodic or quasiperiodic Kitaev chain defined by Eq.~(\ref{Cond_single}).
Now, we remark that the gap size for Bogoliubov quasiparticles depends on the average of the pair potential \cite{Kitaev_stack}.
Thus we assume that the average $c = \sum_{i=1}^{N}\Delta_{i,i+1}/L_{N}$ conserves, where $L_{N}$ is the total bond number of $N$ site chain.
For the $n$th generation of Fibonacci quasicrystals, the bond number $L_{N}$ is represented by $n$th Fibonacci number $F_n$ as $L_{N}=F_{n}$.
Considering the inflation rule, it can be found that $n$th generation of Fibonacci quasicrystals has $F_{n-1}$ and $F_{n-2}$ long and short tiles, respectively.
Therefore, we define the value of the gap function and hopping parameter for short tiles as
\begin{align}
     t_{\rm S} &= \frac{c F_{n}}{F_{n-2}+\rho F_{n-1}},
     \notag\\ 
     \Delta_{\rm S} &= \alpha t_{\rm S}, \quad 0 \leq \alpha \leq 1
    \label{def_param1}
\end{align}
where the parameter $\rho$ is defined as the ratio between the hopping of long and short tiles, denoted as $t_{\rm L}$ and $t_{\rm S}$, $\rho = t_{\rm L}/t_{\rm S}$.
Thus for the periodic and Fibonacci Kitaev chain, $\rho=1$ and $\rho=1/\tau_{\rm g}^2$, respectively, where $\tau_{\rm g}$ is the golden ratio. 
According to this definition, we calculate the differential conductance (Fig.~\ref{Nume_single2_a} and Fig.~\ref{Nume_single2_b}).

As a result, we can see the zero-bias conductance peak.
Therefore, there are Majorana edge modes in both periodic and quasiperiodic cases.
Moreover, within the parameters considered here, quasiperiodicity does not significantly alter the zero-bias conductance peak.
Additionally, we cannot see the effect of potential barrier $U_{\rm b}$.
This could be because the mismatch of the hopping parameter $t_{\rm N}$ and $t_{\rm L}$ between the lead and Kitaev chain works as the potential barrier.

%% file: Appendix/Formulation.tex
In this section, we derive the formulation of the differential conductance for the coupled Kitaev chains.
We consider the model invented by the Ref.~\cite{Kitaev_stack}.
The Hamiltonian of this model is written as
\begin{align}
    H_{S/N} &= \sum_{i=1}^{N}\LB \mu^{\rm S/N} d_{i}^{\rm S/N,\dag} d_{i}^{\rm S/N}\RB 
    \notag\\
    +&\sum_{i=1}^{N-1}\LB -t^{\rm S/N}_{i,i+1} d_{i+1}^{S/N\dag} d_{i}^{\rm S/N} +  h.c. \RB
    \notag\\
    +&\sum_{i=1}^{N-1}\LB \Delta^{\rm S/N}_{i,i+1}d_{i}^{\rm S/N}d_{i+1}^{\rm S/N}  + h.c. \RB,
\end{align}
\begin{align}
    H_{\rm int} &=
    \sum_{i=1}^{N}\LB
    -\tau_{\rm int} d_{i}^{S\dag}d_{i}^{\rm N} 
    + \delta_{\rm int} d_{i}^{\rm S}d_{i}^{\rm N} + h.c.
    \RB,
\end{align}
where the indices S and N represent the superconducting and normal Kitaev chains, respectively.
The parameters $\mu^{\alpha}$, $t^{\alpha}_{i,j}$ and $\Delta^{\alpha}_{i,j}$ are the chemical potential, hopping amplitude and superconducting pair potential of the superconducting ($\alpha=\rm{S}$) or normal ($\alpha=\rm{N}$) chain, respectively.
The parameters $\tau_{\rm int}$ and $\delta_{\rm int}$ are the hopping and pair potential between the superconducting and normal chains.
The operators $d_{i}^{\rm S/N}$ and $d_{i}^{\mathrm{S/N }\dag}$ are the annihilation and creation operators of the superconducting or normal chain at the $i$-th site, respectively.
In the matrix form, the Hamiltonian is written as
\begin{equation}
    \begin{split}
        H_{\rm{couple}} =& \mqty(d^{S\dag} & d^{\rm S} & d^{N\dag} & d^{\rm N} ) \widetilde{H}_{\rm{couple}} \mqty(d^{\rm S} \\ d^{S\dag} \\ d^{\rm N} \\ d^{N\dag} ) ,
        \\
        \check{H}_{\rm{couple}} =&
        \mqty(
        \widetilde{H}_{\rm S}                 &  \widetilde{H}_{\rm{int}} \\
        \widetilde{H}_{\rm{int}}^{\dag} &  \widetilde{H}_{\rm N}
        ),
        \\
        \widetilde{H}_{S/N} =& \mqty(
        \widehat{\mu}^{\rm S/N}              & \widehat{\beta}_{\Delta}^{\rm S/N} &                        & \\
        \widehat{\beta}_{\Delta}^{S/N\dag} & \ddots            & \ddots                 & \\
                               & \ddots            & \ddots                 & \widehat{\beta}_{\Delta}^{\rm S/N} \\
                               &                   & \widehat{\beta}_{\Delta}^{S/N\dag} & \widehat{\mu}^{\rm S/N}  
        ) ,
        \\
        \widehat{\beta}_{\Delta}^{\rm S/N} =& \mqty(
        -t_{i,i+1}^{\rm S/N}        & -\Delta_{i,i+1}^{\rm S/N} \\
        \Delta_{i,i+1}^{\rm S/N}    & t_{i,i+1}^{\rm S/N}
        ),
        \\
        \widetilde{H}_{\rm{int}} =& \mqty(
        -t_{\rm int} I_{\rm N}  &  -\Delta_{\rm int} I_{\rm N} \\
        \Delta_{\rm int} I_{\rm N} & t_{\rm int} I_{\rm N}
        ).\label{eq_app_H_mat}
    \end{split}
\end{equation}
According to the third equation of Eq.~(\ref{eq_app_H_mat}), the self-energies due to the connection with the leads are written as
\begin{align}
    \check{\Sigma}^{\rm S}_{L/R} =\mqty(
    \widetilde{\Sigma}_{L/R} & \widetilde{O}_{2N} \\
    \widetilde{O}_{2N}       & \widetilde{O}_{2N}
    ),
    \quad
    \check{\Sigma}^{\rm N}_{L/R} =\mqty(
    \widetilde{O}_{2N} & \widetilde{O}_{2N} \\
    \widetilde{O}_{2N} & \widetilde{\Sigma}_{L/R}
    )
\end{align}
where $\widetilde{\Sigma}_{L/R}$ is the self-energy due to the connection with the left (right) lead defined in Eq.~(\ref{self_ene_4N}).
Using these definitions, the Green's function is written as
\begin{align}
    \check{G} \equiv \LB z\check{I}_{4N} - \check{H}_{\rm{couple}} -\check{\Sigma}^{\rm \beta}_{\alpha} -\check{\Sigma}^{\rm \beta'}_{\alpha'} \RB^{-1},
\end{align}
where the indices $\alpha = \rm L,$ or $\rm R$ and $\rm\beta = \rm S,$ or $\rm N$ denote the left or right edges of superconducting or normal Kitaev chain.
Here, we consider the current at the $\alpha$ edge of $\beta$ Kitaev chian defined as $I_{\beta_\alpha} = -e\ev{\dot{N}_{\beta_\alpha}}$,
where $\dot{N}_{\beta_\alpha}$ is the time derivative of the number operator at the $\alpha$ edge of $\beta$ Kitaev chain.
To derive the differential conductance depending on the lead connecting patterns, we must consider the tunnel Hamiltonian at the $\alpha$ and $\alpha'$ edges of the $\beta$ and $\beta'$ Kitaev chains.
Thus, the tunnel Hamiltonian is written as
\begin{align}
    H_{\rm T} = t_{\rm tunel} \LB c_{\alpha,0}^{\dag} d_{\alpha}^{\beta} + c_{\alpha',0}^{\dag} d_{\alpha'}^{\beta'} + \mathrm{h.c.} \RB,
\end{align}
where $c_{\alpha,0}^{\dag}$ and $c_{\alpha',0}^{\dag}$ are the creation operators of the lead connected to the $\alpha$ and $\alpha'$ edges of the Kitaev chain, respectively.
In the numerical calculations in the main text, we set the tunneling amplitude as $t_{\rm tunel} = 1.0$ as the unit of energy.
According to this tunneling Hamiltonian, the current at the $\alpha$ edge of $\beta$ Kitaev chain is written as
\begin{align}
    I_{\beta_{\alpha}} = -\frac{ie}{\hbar} \LB \ev{d_{\alpha}^{\beta\dag} c_{\alpha,0}} - \ev{c_{\alpha,0}^\dag d_{\alpha}^{\beta}}  \RB.
\end{align}
Performing the perturbation expansion with respect to the tunneling Hamiltonian, the current is written as
\begin{align}
    &I_{\beta_{\alpha}}
    \notag\\
     &= \frac{ie}{2}\int \frac{d\omega}{2\pi}
    \Tr \LB (I_{\rm N}\otimes\tau_z) \check{\varGamma}_{\beta_{\alpha}}\lB \check{G}^{<} + \check{F}_{\beta_{\alpha}}(\check{G}^r - \check{G}^a) \rB \RB,
    \\
    &\check{\varGamma}_{\beta_\al} = -2\Im \check{\Sigma}_{\al}^{\beta,r},
    \\
    &\check{F}_{\beta_\al} = I_{\rm N} \otimes \mqty(
    f(\hbar\omega-eV_{\beta_\al}) & 0 \\
    0                   & f(\hbar\omega+eV_{\beta_\al})
    ),
    \\
    &\check{\Sigma}_{\al}^{\beta,<} = \check{\varGamma}_{\beta_\al} \check{F}_{\beta_\al},
    \\
    &\check{G}^{<} = \check{G}^r (\check{\Sigma}_{\alpha}^{\beta,<}+\check{\Sigma}_{\alpha'}^{\beta',<}) \check{G}^a,
\end{align}
where the indices $r$, $a$, and $<$ of Green's function and self-energy denote the retarded, advanced, and lesser components, respectively.
Same as the case of a single Kitaev chain, this current can be written as
\begin{align}
    I_{\beta_\alpha}=& e\int \frac{d\omega}{2\pi}
    \left[
    \varGamma_{\beta_{\alpha}}^{-}(\omega)\varGamma_{\beta_{\alpha}}^{+}(\omega)\abs{\check{G}^{r}_{\beta_{\alpha}^{+},\beta_{\alpha}^{-}}(\omega)}^2 \lB f_{\beta_{\alpha}}^{-}-f_{\beta_{\alpha}}^{+} \rB
    \right. 
    \notag \\
    &+ \varGamma_{\beta_{\alpha}}^{-}(\omega)\varGamma_{\beta'_{\alpha'}}^{-}(\omega)\abs{\check{G}^{r}_{\beta_{\alpha}^{+},\beta_{\alpha'}^{\prime +}}(\omega)}^2 \lB f_{\beta_{\alpha}}^{-}-f_{\beta'_{\alpha'}}^{-} \rB
    \notag \\
    &+\left. \varGamma_{\beta_{\alpha}}^{-}(\omega)\varGamma_{\beta'_{\alpha'}}^{+}(\omega)\abs{\check{G}^{r}_{\beta_{\alpha}^{+},\beta_{\alpha'}^{\prime-}}(\omega)}^2 \lB f_{\beta_{\alpha}}^{-}-f_{\beta'_{\alpha'}}^{+} \rB
    \right],
\end{align}
where the indices $\pm$ denote the particle and hole components, respectively.
Therefore, the indices of the Green's function are defined as 
\begin{align}
    \beta_{\alpha}^{\pm}
    = 2\alpha + \beta - \lb \frac{1}{2}\pm \frac{1}{2} \rb,
\end{align}
where $\alpha= 1$, or $N$ for the left or right edge, and $\beta = 0$, or $2N$ for the superconducting or normal Kitaev chain, respectively.
Derivating by the bias voltage, the differential conductance is written as
\begin{align}
    &\pdv{I_{\beta_\alpha}}{V} 
    \notag \\ 
    &= 
    \frac{e^2}{2h}\sum_{E=\pm V/2} \varGamma_{\beta_{\alpha}}^{-} \lb
    \varGamma_{\beta'_{\alpha'}}^{-} \abs{\check{G}^{r}_{\beta_{\alpha}^{+},\beta_{\alpha'}^{\prime-}}}^2 + \varGamma_{\beta_{\alpha}}^{+}\abs{\check{G}^{r}_{\beta_{\alpha}^{+},\beta_{\alpha}^{-}}}^2
    \rb,
\end{align}
where we set as $V_{\beta_{\alpha}} = - V_{\beta'_{\alpha'}} = V/2$.
Here, to make this differential conductance correspond to the net current, we must consider the current from the left and right edges of the lead.
Therefore, we introduce the average current as $I_{\beta_{\alpha}\beta'_{\alpha'}} = \frac{1}{2}\lb I_{\beta_{\alpha}} + I_{\beta'_{\alpha'}} \rb$.
The differential conductance from the average current is written as
\begin{align}
    &\pdv{I_{\beta_{\alpha}\beta'_{\alpha'}}}{V} 
    \notag \\ 
    &= 
    \frac{e^2}{4h}\sum_{E=\pm V/2} \left[
     \varGamma_{\beta_{\alpha}}^{-} \lb
    \varGamma_{\beta'_{\alpha'}}^{-} \abs{\check{G}^{r}_{\beta_{\alpha}^{+},\beta_{\alpha'}^{\prime-}}}^2 
    + \varGamma_{\beta_{\alpha}}^{+}\abs{\check{G}^{r}_{\beta_{\alpha}^{+},\beta_{\alpha}^{-}}}^2
    \rb,
    \right.
    \notag \\ 
    &\quad +
    \left.
    \varGamma_{\beta'_{\alpha'}}^{-} \lb
    \varGamma_{\beta_{\alpha}}^{-} \abs{\check{G}^{r}_{\beta_{\alpha'}^{\prime+},\beta_{\alpha}^{-}}}^2 
    + \varGamma_{\beta'_{\alpha'}}^{+}\abs{\check{G}^{r}_{\beta_{\alpha'}^{\prime+},\beta_{\alpha'}^{\prime-}}}^2
    \rb
    \right].
    \label{eq_app_dI_dV}
\end{align}

\subsection{\texorpdfstring{$S_{L/R}-S_{R/L}$}{S-L/R to S-R/L} channel}
First of all, we consider the $S_{L/R}-S_{R/L}$ channel.
In this case, the differential conductance can be derived perfectly parallel to that of a single Kitaev chain.
However, due to the inversion breaking, the current in the connecting lead is gained by averaging the current from the left and right edges of the lead. 
Therefore, using the Green's function defined as
\begin{align}
    \check{G} \equiv \LB z\check{I}_{4N} - \check{H}_{\rm{couple}} -\check{\Sigma}^{\rm S}_{L} -\check{\Sigma}^{\rm S}_{R} \RB^{-1},
\end{align}
the differential conductance can be formulated as
\begin{align}
    \pdv{I_{S_{L}S_{R}}}{V} =&
    \frac{e^2}{4h}\sum_{E=\pm V/2}
    \left[ 
    \varGamma_{L}^{-} \lb
     \varGamma_{R}^{-} \abs{\check{G}^{r}_{1,2N-1}}^2 + \varGamma_{L}^{+} \abs{\check{G}^{r}_{1,2}}^2
    \rb
    \right.
    \notag \\
    +&
    \left.
    \varGamma_{R}^{-} \lb
     \varGamma_{L}^{-} \abs{\check{G}^{r}_{2N-1,1}}^2 + \varGamma_{R}^{+} \abs{\check{G}^{r}_{2N-1,2N}}^2
    \rb
    \right],
    \label{cond_SS}
\end{align}
where the $\varGamma_{\alpha}^{\pm}$ is the particle and hole components of the imaginary part of the self-energy.

\subsection{\texorpdfstring{$N_{L/R}-N_{R/L}$}{N-L/R to N-R/L} channel}
Even in this case, we can formulate the differential conductance by straightforward calculation.
In this situation, we only need to change the index of position as $(1,2N-1) \to (2N+1,4N-1)$ and $(1,2) \to (2N+1,2N+2)$.
Therefore, the differential conductance can be written as
\begin{align}
    &\pdv{I_{N_{L}N_{R}}}{V}
    =
    \frac{e^2}{4h}
    \notag\\
    &\times
    \sum_{E=\pm V/2} 
    \left[
    \varGamma_{L}^{-}
     \left(
     \varGamma_{R}^{-} \abs{\check{G}^{r}_{2N+1,4N-1}}^2 
     +
     \varGamma_{L}^{+} \abs{\check{G}^{r}_{2N+1,2N+2}}^2
    \right)
    \right.
    \notag\\
    &+
    \left. 
    \varGamma_{R}^{-}
     \left(
     \varGamma_{L}^{-} \abs{\check{G}^{r}_{4N-1,2N+1}}^2 
     +
     \varGamma_{R}^{+} \abs{\check{G}^{r}_{4N-1,4N}}^2
    \right)
    \right].
    \label{cond_NN}
\end{align}

\subsection{\texorpdfstring{$S_{L/R}-N_{R/L}$}{S-L/R to N-R/L} channel}
In this situation, the Green function is defined as
\begin{align}
    \check{G} \equiv 
    \LB
    z\check{I}_{4N} - \check{H}_{\rm{couple}} -\check{\Sigma}^{\rm S}_{L/R} -\check{\Sigma}^{\rm N}_{R/L}
    \RB.
\end{align}
To derive the formula of differential conductance,  $S_{L}-N_{R}$ and $S_{R}-N_{L}$ channels must be considered separately.
According to the definition of self energy (\ref{self_ene_4N}), we can find that
\begin{align}
    &\pdv{I_{S_{L}N_{R}}}{V}=
    \frac{e^2}{4h}
    \notag\\
    &\times\sum_{E=\pm V/2}
    \left[
    \varGamma_{L}^{-} 
     \left(
     \varGamma_{R}^{-} \abs{\check{G}^{r}_{1,4N-1}}^2 
     + 
     \varGamma_{L}^{+} \abs{\check{G}^{r}_{1,2}}^2
    \right)
    \right. 
    \notag\\
    &+
    \left. 
    \varGamma_{R}^{-} 
     \left(
     \varGamma_{L}^{-} \abs{\check{G}^{r}_{4N-1,1}}^2 
     + 
     \varGamma_{R}^{+} \abs{\check{G}^{r}_{4N-1,4N}}^2
    \right)
    \right],
\end{align}
\begin{align}
    &\pdv{I_{S_{R}N_{L}}}{V}=
    \frac{e^2}{4h}
    \notag\\ 
    &\times\sum_{E=\pm V/2} 
    \left[ 
    \varGamma_{L}^{-} 
     \left(
     \varGamma_{R}^{-} \abs{\check{G}^{r}_{2N+1,2N-1}}^2 
     +
     \varGamma_{L}^{+} \abs{\check{G}^{r}_{2N+1,2N+2}}^2
    \right)
    \right. 
    \notag\\
    &+
    \left.
    \varGamma_{R}^{-} 
     \left(
     \varGamma_{L}^{-} \abs{\check{G}^{r}_{2N-1,2N+1}}^2 
     +
     \varGamma_{R}^{+} \abs{\check{G}^{r}_{2N-1,2N}}^2
    \right)
    \right] .
\end{align}

Here, we can get the differential conductance for arbitrary patterns of lead connections.
In the following, we show the detailed representations for our calculations.

\subsection{\texorpdfstring{$S_{L/R}-N_{L/R}$}{S-L/R to N-L/R} channel}
In this case, we can likely formulate the case of $S_{L/R}-N_{R/L}$ channel.
Therefore, we reach the differential conductance with the final situation as
\begin{align}
    &\pdv{I_{S_{L}N_{L}}}{V}=
    \frac{e^2}{4h}
    \notag \\ 
    &\times \sum_{E=\pm V/2}
    \left[ \varGamma_{L}^{-} 
     \left(
     \varGamma_{R}^{-} \abs{\check{G}^{r}_{1,2N+1}}^2 
     + 
     \varGamma_{L}^{+} \abs{\check{G}^{r}_{1,2}}^2
    \right)
    \right.
    \notag\\ 
    &+ \left.
    \varGamma_{R}^{-} 
     \left(
     \varGamma_{L}^{-} \abs{\check{G}^{r}_{2N+1,1}}^2 
     + 
     \varGamma_{R}^{+} \abs{\check{G}^{r}_{2N+1,2N+2}}^2
    \right)
    \right],
\end{align}
\begin{align}
    &\pdv{I_{S_{R}N_{R}}}{V}=
    \frac{e^2}{4h}
    \notag\\ 
    &\times \sum_{E=\pm V/2}
    \left[\varGamma_{L}^{-} 
     \left(
     \varGamma_{R}^{-} \abs{\check{G}^{r}_{2N-1,4N-1}}^2 
     +
     \varGamma_{L}^{+} \abs{\check{G}^{r}_{2N-1,2N}}^2
    \right)
    \right. 
    \notag\\ 
    &+
    \left. 
    \varGamma_{R}^{-} 
     \left(
     \varGamma_{L}^{-} \abs{\check{G}^{r}_{4N-1,2N-1}}^2 
     +
     \varGamma_{R}^{+} \abs{\check{G}^{r}_{4N-1,4N}}^2
    \right)
    \right].
\end{align}